%% file: LASCAS.tex
\documentclass[10pt, conference]{IEEEtran}
\usepackage[moderate,tracking=normal]{savetrees}
\usepackage{pgfplots}
\DeclareUnicodeCharacter{2212}{−}
\usepgfplotslibrary{groupplots,dateplot}
\usetikzlibrary{patterns,shapes.arrows}
\pgfplotsset{compat=newest}
\usepackage{cite}
\usepackage{amsmath,amssymb,amsfonts}
\usepackage{algorithmic}
\usepackage{graphicx}
\usepackage{textcomp}
\usepackage{xcolor}
\usepackage{cite}
\usepackage{amsmath,amssymb,amsfonts}
\usepackage{algorithmic}
\usepackage{graphicx}
\usepackage{textcomp}
\usepackage{xcolor}
\usepackage{cite}
\usepackage{epsfig}
\usepackage{fancyhdr}
\usepackage{color}
\usepackage{graphicx}
\usepackage[linesnumbered,ruled,vlined]{algorithm2e}
\usepackage{amsmath}
\usepackage{algorithmic}
\usepackage[T1]{fontenc}
\usepackage{times}
\usepackage{epsfig}
\usepackage{fancyhdr}
\usepackage{framed}
\usepackage{multirow}
\usepackage{url}
\usepackage{balance}
\usepackage{booktabs}
 \usepackage{graphicx}
\usepackage{listings}
\usepackage{caption}
\usepackage{upquote}
\usepackage{xcolor}
\usepackage{comment}
\usepackage{subfigure}
\usepackage{mathtools}
\usepackage{ragged2e}
\usepackage{lastpage}
\usepackage{academicons}

\usepackage{hyperref}
\usepackage{array}
\usepackage{epsfig}
\usepackage{fancyhdr}
\usepackage{color}
\usepackage[linesnumbered,ruled,vlined]{algorithm2e}
\usepackage{amsmath}
\usepackage{times}
\usepackage{framed}
\usepackage{multirow}
\usepackage{url}
\usepackage{balance}
\usepackage{booktabs}
\usepackage{listings}
\usepackage{caption}
\usepackage{upquote}
\usepackage{comment}
\usepackage{subfigure}
\usepackage{mathtools}
\usepackage{ragged2e}
\usepackage{lastpage}
\usepackage{academicons}
\usepackage{adjustbox}
\usepackage{hyperref}
\usepackage{adjustbox}
\usepackage{academicons}
\usepackage{hyperref}
\usepackage{tabularx}
\newcommand*\CircledText[1]{\tikz[baseline=(char.base)]{
            \node[shape=circle,draw,inner sep=0.5pt] (char) {#1};}}
\renewcommand{\thefootnote}{\textit{\alph{footnote}}}

\ifCLASSINFOpdf
\else
\fi
\begin{document}
\newcommand{\papername}{Ellora}
\title{\papername{}: \texorpdfstring{\underline{E}xp\underline{l}oring \underline{L}ow-Power \underline{O}FDM-based \underline{R}adar Processors using \underline{A}pproximate Computing}{Ea}}

\author{
  \IEEEauthorblockN{Rajat Bhattacharjya\IEEEauthorrefmark{1}, Alish Kanani\IEEEauthorrefmark{2}, A Anil Kumar\IEEEauthorrefmark{3}, Manoj Nambiar\IEEEauthorrefmark{3}, M Girish Chandra\IEEEauthorrefmark{3}, and Rekha Singhal\IEEEauthorrefmark{4}
   }
   \IEEEauthorblockA{\IEEEauthorrefmark{1}University of California, Irvine}
  \IEEEauthorblockA { \IEEEauthorrefmark{2}University of Wisconsin-Madison}
  \IEEEauthorblockA { \IEEEauthorrefmark{3}TCS Research, India}
   \IEEEauthorblockA {\IEEEauthorrefmark{4}TCS Research, USA\\
   rajatb1@uci.edu, ahkanani@wisc.edu, \{achannaanil.kumar, m.nambiar, m.gchandra, rekha.singhal\}@tcs.com }}

%\newcommand{\short}{ACLA }
%\newcommand{\shortx}{ACLA}



%\author{\IEEEauthorblockN{Rajat Bhattacharjya\IEEEauthorrefmark{1},
%Telephone: (800) 555--1212, Fax: (888) 555--1212}


\makeatletter
\def\footnoterule{\kern-3\p@
  \hrule \@width 2in \kern 2.6\p@} % the \hrule is .4pt high
\makeatother
\newcommand{\copyrightnotice}[1]{{%
  \renewcommand{\thefootnote}{}% Remove footnote number
  \footnotetext[0]{#1}%
}}

\maketitle
\copyrightnotice{
\vspace{-10pt}

\noindent Paper accepted at IEEE-LASCAS 2024. 

\noindent Authors’ version of the work posted for personal use and not for redistribution. The definitive version will be available in the Proceedings of the IEEE-LASCAS 2024.}
% \IEEEoverridecommandlockouts
% \IEEEpubid{\makebox[\columnwidth]
% {979-8-3503-8122-1/24/\$31.00~\copyright2024 IEEE \hfill}
% \hspace{\columnsep}\makebox[\columnwidth]{ }}
% \IEEEpubidadjcol
\begin{abstract}
In recent times, orthogonal frequency-division multiplexing (OFDM)-based radar has gained wide acceptance given its applicability in joint radar-communication systems. 
However, realizing such a system on hardware poses a huge area and power bottleneck given its complexity. Therefore it has become ever-important to explore low-power OFDM-based radar processors in order to realize energy-efficient joint radar-communication systems targeting edge devices. This paper aims to address the aforementioned challenges by exploiting approximations on hardware for early design space exploration (DSE) of trade-offs between accuracy, area and power. We present Ellora, a DSE framework for incorporating approximations in an OFDM radar processing pipeline. Ellora uses pairs of approximate adders and multipliers to explore design points realizing energy-efficient radar processors. Particularly, we incorporate approximations into the block involving periodogram based estimation and report area, power and accuracy levels. Experimental results show that at an average accuracy loss of 0.063\% in the positive SNR region, we save 22.9\% of on-chip area and 26.2\% of power. Towards achieving the area and power statistics, we design a fully parallel Inverse Fast Fourier Transform (IFFT) core which acts as a part of periodogram based estimation and approximate the addition and multiplication operations in it. The aforementioned results show that Ellora can be used in an integrated way with various other optimization methods for generating low-power and energy-efficient radar processors.
\end{abstract}



% \fi
\IEEEpeerreviewmaketitle

\input{Text/intro}

%\input{Text/background}
\input{Text/method}
\input{Text/evaluation.tex}
% \input{Text/analysis}
%\input{Text/related}
\input{Text/conclusion}
\input{LASCAS.bbl}
\end{document}

%% file: Text/intro.tex
\section{Introduction}
\label{sec:intro}
%Context: what is the system/application, why do I care?
%Challenge: what’s so hard? What’s the problem to solve? 
%Opportunity- gap in existing work?
%Contribution: how are you going to fill the gap in solving the problem? What do you bring that’s novel?
Systems involving sensing and communication have been historically separate~\cite{jrc}. However, recently the domain of joint radar-communication (JRC) has emerged, where a radar and a communication system are co-located in a single system. Such a system helps in optimal utilization of spectral resources and benefits both sensing and signalling via cooperation between them~\cite{jrcx}. JRC systems find applications in vehicle-to-vehicle (V2V) communication scenarios for enabling various safety functions, smart traffic applications and developing autonomous vehicles~\cite{jrc1}. However such systems suffer from great complexity and hardware overheads given the complex digital signal processing techniques employed in them~\cite{jrc2}. Systems involving orthogonal frequency-division multiplexing (OFDM) waveforms offer great interoperability between both radar and communication system~\cite{jrc}. However, in order to target power-constrained edge devices, it becomes important to achieve solutions targeting low-power and energy-efficient hardware.

There have been some works targeting efficient hardware in JRC. One work focused on introducing zero-padding in OFDM-based radar as opposed to the usual cyclic prefix~\cite{zeropad}. %as it doesn't require full-duplex, given there is no self-interference (SI) in the guard interval~\cite{zeropad}.
Another work focused on introducing hybrid precoding and dynamic selection of optimal RF chains~\cite{rfc}. Kaushik et al. proposed an energy-efficient RF chain and DAC bit selection procedure~\cite{rfc2}. However studies focusing on low-power and energy-efficient systems in particular for JRC are limited and thus in this paper, we focus on addressing this problem.
%However, studies focusing on low-power and energy-efficient JRC systems have been quite minimal which provides us with an opportunity to explore more solutions. 

%In this study, we proposed a ZP-OFDM-JC&S system tosimplify the hardware design at the receiver. Unlike with CPOFDM radar, ZP-OFDM radar does not need full duplex, asthere is no self-interference during the GI, which can be usefor radar detection. We proposed a signal-processing algorithmfor radar detection in that case, and showed that when the selfinterference in CP-OFDM is high, the ZP-OFDM is beneficial.This advantage depends on the target range, since this willdetermine the amount of gathered energy in the GI. We showthen that ZP-OFDM can be a cost-effective and efficientalternative to CP-OFDM when the self-interference isolationis low.

\begin{figure}
\begin{center}
    \includegraphics[width=0.4\textwidth]{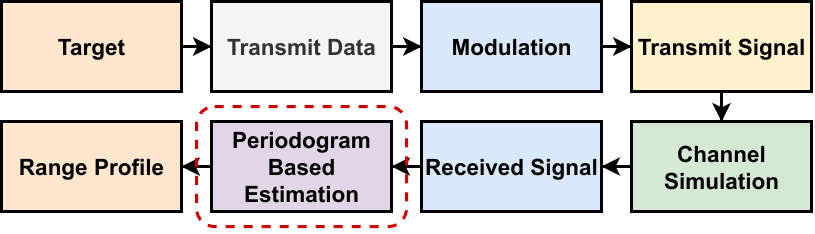}
    \caption{OFDM Radar processing pipeline. Approximated block highlighted in red.}
    \label{fig:outline}
    \end{center}
    \vspace{-5ex}
\end{figure}

In this paper, we propose a design space exploration (DSE) framework, \textit{Ellora} for generating low-power radar processors pertaining to OFDM waveforms using approximate computing~\cite{approx}. Approximate computing is a computing paradigm that aims at achieving energy efficiency at the expense of tolerable inexactness in applications~\cite{acla, approx, lascas23error}. Arithmetic units such as adders and multipliers form the basic building blocks in all such approximate systems~\cite{locate, cesa, bio1, rao}. Ellora helps find out the approximation target in the application pipeline and incorporates pairs of approximate adders and multipliers in the targeted block to achieve area and power savings with just a  marginal loss of accuracy. The approximation target is found to be the block involving periodogram based estimation (shown in Fig.~\ref{fig:outline}) since it involves the computationally-intensive Inverse Fast Fourier Transform (IFFT) and shows error-resilience. Though there have been some works focusing on approximate FFT~\cite{fft1,fft2}, we showcase a DSE framework across an end-to-end application pipeline (OFDM-radar for JRC) along with a fully parallel hardware design. Ellora explores optimal accuracy-power-area design points in order to generate low-power radar processors.
%Particularly, Ellora approximates the computationally-intensive and error-resilient Inverse Fast Fourier Transform (IFFT) operation~\cite{ifftex} in the periodogram based estimation block (the approximation target) as shown in Fig.~\ref{fig:outline} and explores pareto-optimal accuracy-power-area design points. 
Experiments show that at an average accuracy loss of 0.063\% in the positive SNR region, Ellora helps save 22.9\% of on-chip area and 26.2\% of power.

%points
%jrc
%approx- already too much focus on ai ml, too much scope, behaviours can be different than expected, cant generalize, application exploration

%% file: Text/method.tex
\section{Methodology}

\begin{figure*}
    \includegraphics[width=1\textwidth]{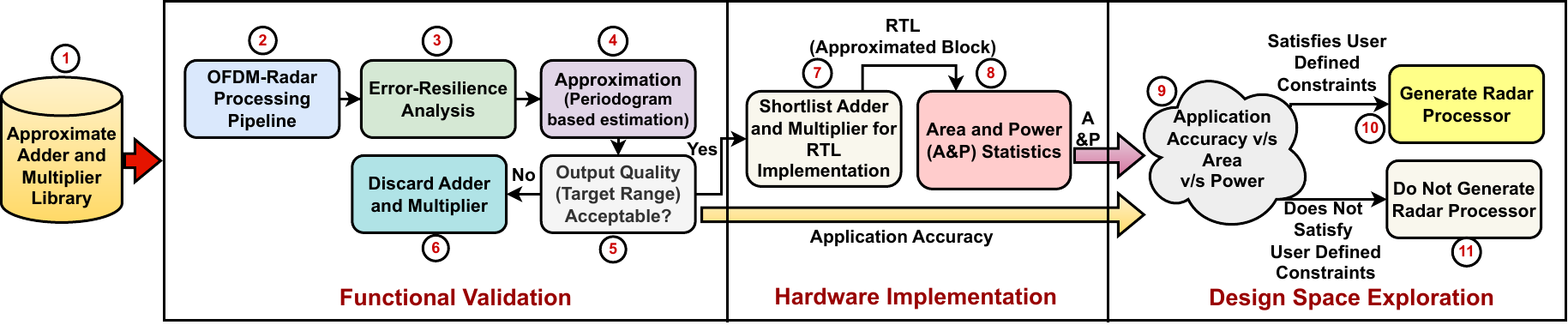}
    \caption{Methodology of Ellora}
    \label{fig:radarProcessing}
    \vspace{-2ex}
\end{figure*}
\label{sec:method}
Fig.~\ref{fig:outline} outlines the complete radar processing pipeline and highlights the block of approximation, i.e, the block involving periodogram based estimation. Towards achieving the goal of generating low-power OFDM-based radar processors, we follow the flow as shown in Fig. ~\ref{fig:radarProcessing}. The process of generating low-power OFDM radar processors (\textit{Ellora}) involves 3 steps mainly: Functional Validation at the Software Level, Hardware Implementation, followed by Design Space Exploration to generate and explore optimal design points. The steps are described in continuation, where the circled numbers denote the blocks in Fig. ~\ref{fig:radarProcessing}:

\begin{comment}
\begin{figure}
    \includegraphics[width=0.5\textwidth]{Images/butterfly.pdf}
    \caption{Approximating the butterfly unit: Approximated operators highlighted in red}
    \label{fig:butterfly}
\end{figure}
\end{comment}
\subsection{Functional Validation}We conduct Functional Validation (as shown in Fig.~\ref{fig:radarProcessing}) at the Software Level using MATLAB so as to study the effects of approximation on the end application, i.e., target detection in OFDM radar. 

First, we take an OFDM-radar processing pipeline in consideration\textsuperscript{\CircledText{2}} (shown in Fig.~\ref{fig:outline}). Then we conduct the error-resilience analysis\textsuperscript{\CircledText{3}} for the whole pipeline by injecting white Gaussian noise along with respective inputs into the computationally-intensive digital signal processing blocks. This helps us investigate which blocks are resilient enough to handle the effects of approximations and thereby we select an approximation target. We select the block involving periodogram based estimation as the approximation target since it is computationally-intensive~\cite {periodo} and shows error resilience. %We say that the block involving periodogram based estimation shows error resilience because inspite of some inexactness in processed data upon introducing WGN, it produces results of acceptable quality. 
The periodogram based estimation that we employ here eventually performs IFFT on the data obtained after the element-wise division of received data by transmitted data, resulting in normalized power and finally helping plot range of target (m) v/s normalized range profile (dB).

Then we approximate\textsuperscript{\CircledText{4}} the addition and multiplication operations in the periodogram based estimation block (in the IFFT) by employing an adder-multiplier\textsuperscript{\CircledText{1}} pair at once from the EvoApprox Library~\cite{evoapprox}. The selection of adder-multiplier pairs is done based on their individual error metrics~\cite{evoapprox} and various combinations are tried out keeping in mind corner cases (best and worst individual error metrics) so that a representative design space can be explored. The IFFT operation that we perform is a radix-2 decimation in time. We employ approximate adders and multipliers in the IFFT core. Since now we have an approximated IFFT core, thereby an approximated OFDM radar processing pipeline, we obtain the target's range profile. If we find that the deviation of target's range is not at an acceptable level, we discard the adder and multiplier\textsuperscript{\CircledText{6}} pair in consideration. Else we move to the hardware implementation part where we design and implement our IFFT core and introduce hardware models of the approximate adder-multiplier pairs. 

The use of approximate circuits can have varying impacts on the accuracy of an application. Therefore, within this methodology, it is crucial to conduct a comprehensive exploration of the design space to identify suitable circuits for the application and understand the underlying reasons. Although circuits may possess diverse accuracy metrics such as Error Percentage (EP), Mean Absolute Error (MAE), Worst-Case Absolute Error (WCE), Mean Relative Error (MRE) etc. as defined in~\cite{stat}, it is essential to examine their end effects on applications and end-to-end pipelines, as the outcomes may differ from what is expected.
%It is important to note that approximate circuits have varying impacts on an application's accuracy. Therefore, with such a methodology, it is crucial to conduct a design space exploration so as to figure out which circuits are suitable for the application and why. Though circuits might have various accuracy metrics associated with themselves~\cite{stat}, it is essential for us to study the end effects of them on applications, and end-to-end pipelines as the results might differ from what is expected.

\subsection{Hardware Implementation}
\label{sec:hw}

\begin{figure}
    \includegraphics[width=0.48\textwidth]{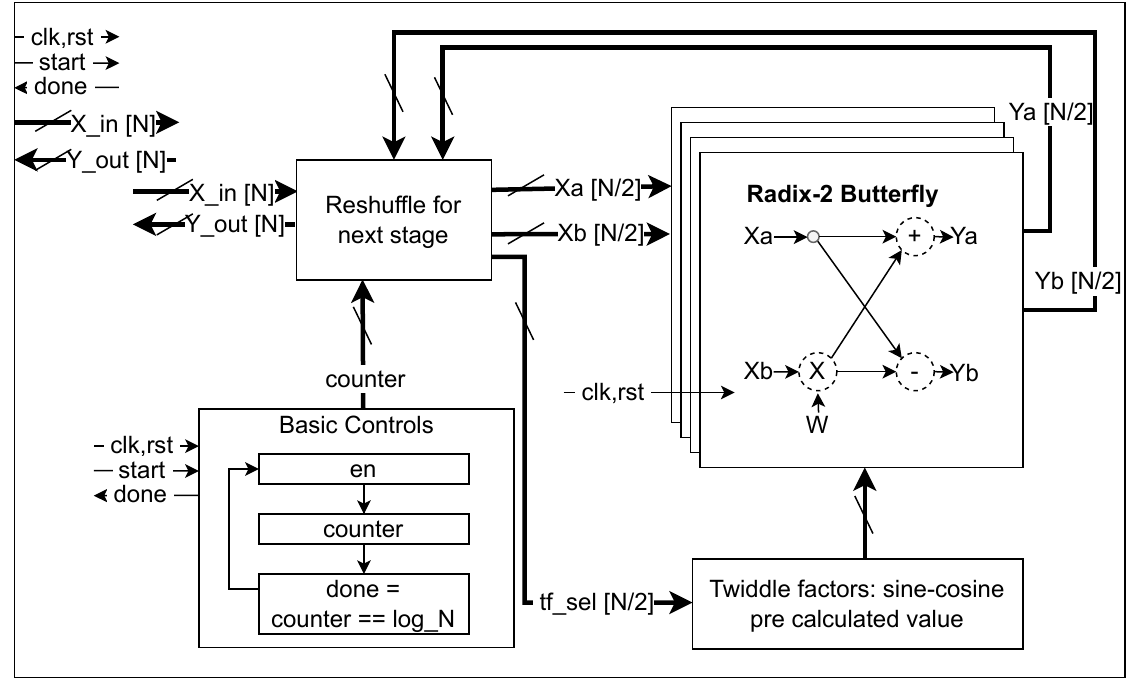}
    \caption{IFFT Core Microarchitecture}
    \label{fig:ifft}
    \vspace{-3ex}
\end{figure}
%Once we have shortlisted adders and multipliers\textsuperscript{\CircledText{7}} from the Functional Validation part, we proceed to incorporate those circuits into our IFFT core and obtain area and power statistics\textsuperscript{\CircledText{8}}. The IFFT core was described using SystemVerilog and synthesized with using 45nm NanGate Open Cell Library with 100 MHz frequency
For Hardware Implementation, first, we design our own IFFT core and describe it using SystemVerilog. Its microarchitecture is shown in Fig. ~\ref{fig:ifft}. It is a fully parallel IFFT core which calculates IFFT for a given sequence of N-points in $log_2(N)$ clock cycles.

The central computation unit of this IFFT core is the radix-2 butterfly structure. This module calculates the complex multiply-accumulation in a single cycle which is described in the Eq.~\ref{eq_bfly}, where $Xa$ and $Xb$ are complex inputs while $W$ is the \textit{twiddle factor}. For an N-point IFFT, $W_N$ is $e^{-i{2\pi/N}}$ which can also be represented in terms of $sine$ and $cosine$ values.
%for this study we used 512 point IFFT to align with functional validation. This means the core is capable of doing each stage of IFFT in a single cycle.

\begin{equation} \label{eq_bfly}
\begin{split}
    Ya = Xa + W*Xb \\
    Yb = Xa - W*Xb
\end{split}
\end{equation}

This complex multiply-accumulate operation (as shown in Eq.~\ref{eq_bfly}) uses 4 multipliers and 6 adders. The output $Ya$ and $Yb$ are being stored in flip-flops which are then used as inputs to the next IFFT stage. Since all the butterfly units for a particular IFFT stage are independent, all can be calculated parallelly. In an N-point IFFT calculation, each stage consists of N/2 butterfly structures. The output of N/2 butterfly units are then fed back to the reshuffle module, which reshuffles $Ya$ and $Yb$ as inputs to the next stage and selects the \textit{twiddle factor} using a counter which counts till $log_2(N)$. All possible $sine$ and $cosine$ values (\textit{twiddle factors}) are pre-calculated and stored in a Read-Only Memory (ROM). When the counter reaches $log_2(N)$, outputs of N/2 butterfly units are combined as final output. 

The butterfly module uses 16-bit signed adder and multiplier units which are approximated~\cite{evoapprox} in this work. This is to align with the functional validation as 16-bit signed numbers were used for OFDM-radar processing pipeline. The approximated adders and multipliers are highlighted as broken circles in the microarchitecture (Fig.~\ref{fig:ifft}). 

Now, as per the flow in Fig.~\ref{fig:radarProcessing}, once we have shortlisted adders and multipliers\textsuperscript{\CircledText{7}} from the Functional Validation part, we proceed to incorporate those circuits into our IFFT core and obtain area and power statistics\textsuperscript{\CircledText{8}}. Towards achieving the area and power statistics, we synthesize our accurate and approximate IFFT cores using the 45nm NanGate Open Cell Library with 100 MHz frequency in Synopsys Design Compiler (DC). Next, we conduct the design space exploration by varying the adder-multiplier pairs.

%Once we have shortlisted adders and multipliers\textsuperscript{\CircledText{7}} from the Functional Validation part, we proceed to incorporate those circuits into our IFFT core and obtain area and power statistics\textsuperscript{\CircledText{8}}. The IFFT core was described using SystemVerilog and synthesized with using 45nm NanGate Open Cell Library with 100 MHz frequency. Next, we conduct the design space exploration by varying the adder-multiplier pairs.

\subsection{Design Space Exploration}
%Since now we have the statistics such as accuracy, area, and power\textsuperscript{\CircledText{9}}, it becomes important to identify design points satisfying user-defined quality constraints. Thus, we conduct the design space exploration of radar processors so as to obtain optimal design points. 
After functional validation and hardware evaluation, if we find that the trade-off between accuracy, area, and power\textsuperscript{\CircledText{9}} is satisfactory, the adder-multiplier pair corresponding to the design point is used to generate a radar processor\textsuperscript{\CircledText{10}}. In case the design point fails to satisfy user-defined quality constraints, we discard the adder-multiplier pair corresponding to that design point and do not generate radar processor\textsuperscript{\CircledText{11}}. %It is worth noting that the 
The selection in \CircledText{6} is different from the one in \CircledText{11}. \CircledText{6} enables early selection just after functional validation at the software level, but, in order to reach \CircledText{11}, it is imperative to carry out functional validation and hardware implementation, followed by design space exploration. 

%% file: Text/evaluation.tex
\section{Evaluation}
\label{sec:eval}
This section provides us with the comprehensive evaluation of the flow presented in Fig. ~\ref{fig:radarProcessing}. The system properties for the pipeline presented in Fig.~\ref{fig:outline} is shown in Table~\ref{tab:sys}.
%We present an OFDM-radar pipeline involving various blocks as shown in Fig.~\ref{fig:outline}. The system properties for the pipeline is shown in Table~\ref{tab:sys}. As mentioned in Section~\ref{sec:method}, we follow this flow: Functional Validation -> Hardware Implementation -> Design Space Exploration (DSE). 

%Functional Validation provides us with accuracy statistics for the application in consideration, i.e, OFDM-radar pipeline. For hardware implementation, we design our own IFFT core and approximate the adder-multiplier pairs in the butterfly unit so as to obtain area and power results. Finally, we conduct the DSE which helps us explore design points satisfying user-defined quality constraints corresponding to the parameters of accuracy, area, and power, thereby helping us generate low-power and energy-efficient radar processors.
\begin{table}[]

\caption{System parameters corresponding to Fig.~\ref{fig:outline}}
\label{tab:sys}
\begin{adjustbox}{width=\columnwidth,center}
\begin{tabular}{|l|l|}
\hline
\textbf{System Parameter/Block} & \textbf{Value/Property}               \\ \hline
Carrier frequency               & 30 GHz                                \\ \hline
Number of subcarriers           & 32                                    \\ \hline
Number of symbols               & 16                                    \\ \hline
Subcarrier spacing              & 960 kHz                               \\ \hline
Elementary OFDM symbol duration & 1.04 $\mu$s                          \\ \hline
Cyclic prefix duration          & 0.26 $\mu$s                          \\ \hline
Total symbol duration           & 1.3 $\mu$s                           \\ \hline
Modulation                      & 4-QAM                                 \\ \hline
%Zadoff-Chu sequence index       & 5                                     \\ \hline
%Zadoff-Chu sequence root        & 29                                    \\ \hline
Target position                 & 50 m                                  \\ \hline
Target velocity                 & 20 m/s                                \\ \hline
Channel                         & Additive White Gaussian Noise Channel \\ \hline
\end{tabular}
\end{adjustbox}
\end{table}

%subsection{Experimental Setup}
%\subsubsection{Behavioural/Functional Model}
%\subsubsection{Hardware Model}
\begin{figure}
\begin{center}
    \includegraphics[width=0.38\textwidth]{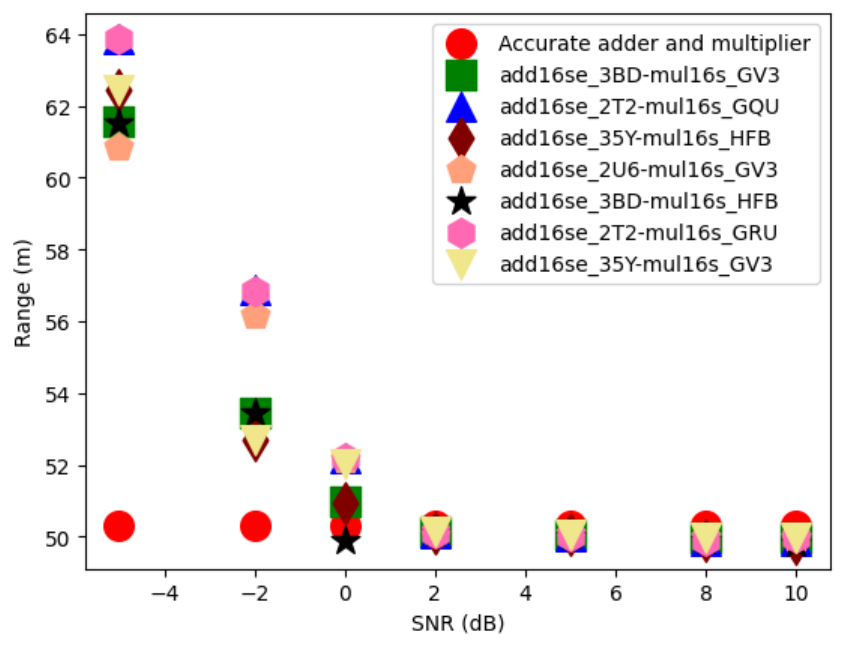}
    \caption{Accuracy statistics of various adder-multiplier (accurate and approximate) pairs. Range averaged for 100 runs per SNR for each adder-multiplier pair.}
    \label{fig:acc}
    \end{center}
    \vspace{-4ex}
\end{figure}

\subsection{Accuracy Analysis}
\label{sec:acc}
For obtaining accuracy statistics, we approximate the addition and multiplication operations in the butterfly unit of the IFFT core (periodogram based estimation) and then obtain results for target's range across an end-to-end pipeline as shown in Fig.~\ref{fig:outline}. We also apply Zadoff-Chu precoding during transmission to improve the correlation properties of the transmitted signal.

\begin{figure}
\begin{center}
    \includegraphics[width=0.38\textwidth]{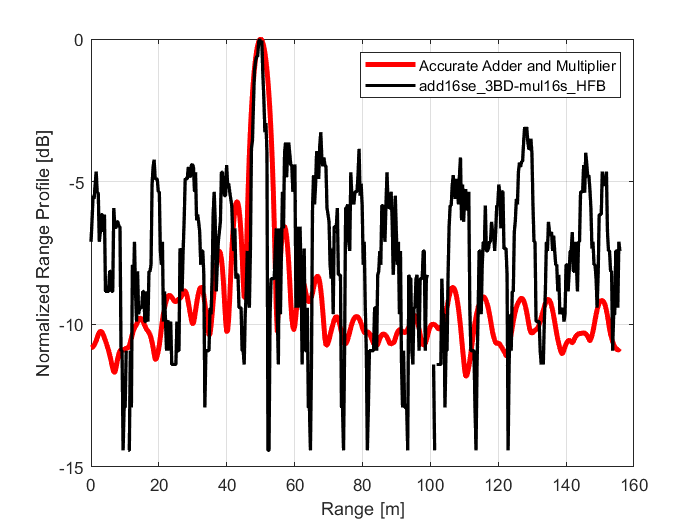}
    \caption{Target's range profile at SNR= 5 dB with accurate adder and multiplier; and add16se\_3BD-mul16s\_HFB}
    \label{fig:periodogram}
    \end{center}
    \vspace{-2ex}
\end{figure}

Fig.~\ref{fig:acc} shows the value of averaged target's range (original target is at distance = 50 m) over an SNR range of -5 to 10 dB using various adder and multiplier pairs, including accurate and approximate circuits~\cite{evoapprox}. The range values obtained per SNR are averaged across 100 runs. Fig.~\ref{fig:periodogram} shows the target's range profile using accurate adder and multiplier; and the approximate pair add16se\_3BD-mul16s\_HFB. We can see that the highest peaks (estimated target range) for both accurate and approximate circuit pairs lie close to each other.

Thus, we make four major observations. First, pertaining to Fig.~\ref{fig:acc}, results show that in the negative SNR region, the accurate adder and multiplier perform best. Second, we can see that at SNR=0 dB, the pair of add16se\_3BD-mul16s\_HFB performs slightly better (deviation= 0.19\%) than the accurate adder and multiplier pair (deviation= 0.63\%) as the range obtained is closer to 50 m (the target's actual range). Third, for negative SNR region, all approximate circuit pairs have high deviation, and gradually moving towards positive SNR, we see that the approximate circuits tend to give results as accurate ones. Fourth, the pair add16se\_3BD-mul16s\_HFB performs well for estimating the target's range. This is interesting because add16se\_3BD has an MAE of 0.046\%, EP of 99.02\%, MRE of 0.96\% which is the highest among all approximate adders in consideration~\cite{evoapprox}. Similarly, the multiplier mul16s\_HFB has an MAE of 0.002\%, EP of 98.43\%, MRE of 0.22\% which is also quite high~\cite{evoapprox}. However, the overall effect after selecting such a pair seems to give us good results in the positive SNR region. This illustrates why a method like Ellora is necessary. Dynamic interactions between system modules can yield unexpected results due to factors like input data distribution and overall functional design etc.% This is an important case which highlights why such a method as Ellora is necessary as there are many dynamic interactions between various modules across an end-to-end system that might ultimately give results that are different from what is expected. Such cases can be credited to various factors such as input data distribution, the overall functional design etc.

Finally, it is important to note that for practical applications, considering positive SNR region, approximate circuits seem to be a viable option for generating radar processors.

%\begin{figure*}
    
 %  \includegraphics[width=1\textwidth]{Images/periodogram.pdf}
  %  \caption{Target's range profile using various adder-multiplier pairs at SNR= 5 dB.
   % \textbf{First row (Left to right)}: Accurate adder and multiplier, add16se\_3BD-mul16s\_GV3, add16se\_2T2-mul16s\_GQU, add16se\_35Y-mul16s\_HFB.
%\textbf{Second row (Left to right)}: add16se\_2U6-mul16s\_GV3, add16se\_3BD-mul16s\_HFB, add16se\_2T2-mul16s\_GRU, add16se\_35Y-mul16s\_GV3.}
 %   \label{fig:periodogram}
%\end{figure*}

\subsection{Hardware Evaluation}
\begin{figure}
\begin{center}
    \includegraphics[width=0.4\textwidth]{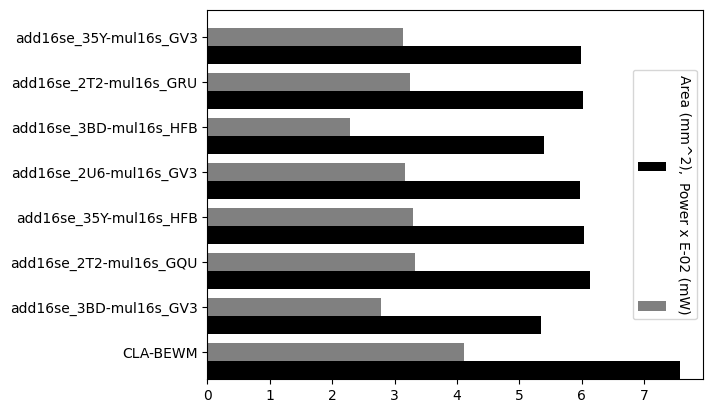}
    \caption{Area and Power Statistics of 512-point IFFT core using various adder-multiplier pairs}
    \label{fig:aps}
    \vspace{-3ex}
    \end{center}
 
\end{figure}

\vspace{-1ex}
As stated in Section~\ref{sec:hw}, we implement our 512- point IFFT core (since, number of subcarriers (32) $\times$ number of symbols (16) = 512) and incorporate various adder-multiplier pairs in it. %Finally, we obtain area and power statistics using DC with a 45nm technology node.
The comprehensive area and power statistics are shown in Fig.~\ref{fig:aps}. We  select Carry-Lookahead Adder (CLA)-Booth Encoded Wallace Multiplier (BEWM) as the accurate circuit pair so as to meet the high frequency requirement of 100 MHz, which is not fulfilled using designs involving long carry chains.  From Fig.~\ref{fig:aps}, we can see that the combination of CLA-BEWM takes up the most area and power. %Approximate circuit pairs easily meet the desired frequency as they lack long carry chains in their design.
 On average, considering all approximate adder-multiplier pairs, the area and power savings compared to the accurate case is 22.9\% and 26.2\% respectively.

The combination of add16se\_3BD-mul16s\_HFB provides the highest power savings, i.e., 44.4\%. It also provides area savings of 28.83\% when compared to the accurate case. This provides us with a very good relationship between accuracy and hardware statistics, as we can see from Section~\ref{sec:acc} that among all approximate circuit pairs, add16se\_3BD-mul16s\_HFB seems to perform best in terms of accuracy. And now in terms of hardware too, it seems the same pair gives out good results. This drives us to explore design options for achieving optimal points while meeting user-defined quality constraints, enabling decisions on low-power radar processors for these points.%This motivates us to explore the design space for obtaining optimal design points satisfying user-defined quality constraints and thereby make decisions on generating low-power radar processors corresponding to such points.
\begin{figure}
\begin{center}
       \includegraphics[width=0.36\textwidth]{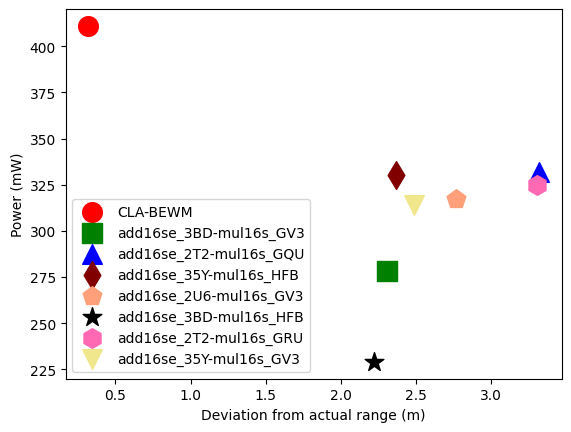}
    \caption{Deviation from actual range (m) v/s Power (mW)}
    \label{fig:pd}
%    \vspace{-4ex}
\end{center}
\vspace{-3ex}
\end{figure}
\begin{figure}
   \begin{center}
   \includegraphics[width=0.36\textwidth]{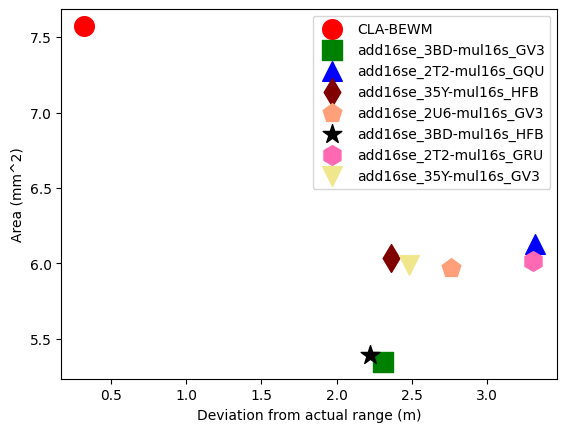}
    \caption{Deviation from actual range (m) v/s Area (mm$^2$)}
    \label{fig:ad}
    \end{center}
\vspace{-3ex}
\end{figure}
\subsection{Design Space Exploration}
%Since now we have both accuracy and hardware statistics, we present a 3D grid so as to explore the design space to generate low-power radar processors as shown in Fig.~\ref{fig:dse}. The 3 axes in the grid are: deviation from actual range (in metres), area (mm$^2$), and power (mW). The deviation from actual range (in metres) is the measure of accuracy for a particular adder-multiplier pair (averaged across SNR= -5 to 10 dB). It shows across an end-to-end pipeline what the target range is after using various adder-multiplier pairs in the periodogram based estimation block (IFFT core). From Fig.~\ref{fig:dse}, we can make various decisions based on user-defined quality constraints. E.g.; for a power budget of < 300 mW, we have 2 pairs of approximate adders and multipliers satisfying it, namely, add16se\_3BD-mul16s\_GV3, and add16se\_3BD-mul16s\_HFB. If we also couple deviation from actual range being < 2.3 m along with a power budget of < 300 mW, then we only obtain the design point corresponding to add16se\_3BD-mul16s\_HFB. Similarly, we can make many observations and report design points pertaining to various accuracy-area-power combinations and thereby generate low-power radar processors.
We study the relationship between accuracy and hardware statistics so as to explore the design space in order to generate low-power radar processors. Fig.~\ref{fig:pd} and~\ref{fig:ad} show the relationship between deviation from the actual range (in metres, averaged across SNR= -5 to 10 dB) and power (mW) and area (mm$^2$) respectively. The deviation from actual range is obtained after using various adder-multiplier pairs in the periodogram based estimation block (IFFT core) across an end-to-end pipeline as shown in Fig.~\ref{fig:outline}. From Fig.~\ref{fig:aps},~\ref{fig:pd},~\ref{fig:ad}, we can make various decisions based on user-defined quality constraints. E.g.; referring to Fig.~\ref{fig:pd}, for a power budget of < 300 mW, we have 2 pairs of approximate adders and multipliers satisfying it, namely, add16se\_3BD-mul16s\_GV3, and add16se\_3BD-mul16s\_HFB. If we also couple deviation from actual range being < 2.3 m along with a power budget of < 300 mW, then we only obtain the design point corresponding to add16se\_3BD-mul16s\_HFB. Similarly, we can make many observations and report design points pertaining to various accuracy-area-power combinations for generating low-power radar processors.
These results apply exclusively to the models and parameters shown in Fig.~\ref{fig:outline},  ~\ref{fig:ifft}, and Table~\ref{tab:sys}. With change in models and parameters, Ellora can help find optimal design points.
  
  % \begin{figure}[tb]
   % \begin{center}
   %         \mbox{ 
            %\hspace{-1.0ex}
           %\subfigure[Deviation from actual range (Accuracy) v/s Power (mW)]
    %        {
     %           \label{fig:area}
      %         \includegraphics[width=0.24\textwidth]{Images/pd.png}
       %         }
            
            %\subfigure[Deviation from actual range (Accuracy) v/s Area (mm$^2$)]
         %      {
        %       \label{fig:power}
          %     \includegraphics[width=0.24\textwidth]{Images/ad.png}
           %     }
           
            %}
            %\caption{Area and Power Statistics of 512-point IFFT core using various adder-multiplier pairs}
           % \vspace{-3.0ex}
            %\label{fig:aps} 
        %\end{center}
  %  \end{figure} 
   

%% file: Text/conclusion.tex
%\vspace{-3ex}
\section{Conclusion and Future Work}
We presented Ellora, a DSE framework (across an end-to-end pipeline) that leverages approximate computing and helps explore optimal design points for generating low-power OFDM-based radar processors. As a part of exploration of design points, we also propose a radix-2 fully parallel IFFT core for various use cases. Ellora exploits pairs of approximate adders and multipliers inside the compute-intensive IFFT core (a part of the periodogram based estimation block) in the OFDM radar processing pipeline. Experimental results show that at a marginal loss of accuracy, Ellora is able to discover optimal design points satisfying user-defined quality constraints while saving both on-chip area and power. 

%It is also seen that Ellora discovers interesting design points that give unexpected results, thereby supporting such a DSE framework since every application exhibits a different level of sensitivity to approximations. 
It is also seen that Ellora discovers interesting design points that give unexpected results, which shows the usefulness of such a DSE framework. This is particularly so since every application exhibits a different level of sensitivity to approximations.
In the future, we plan on integrating Ellora with various other optimization methods, e.g.,~\cite{ofdmr,sen1} and study the effects on end-to-end pipelines as a result of such integration.